\documentclass[reprint,
 amsmath,amssymb,
 aps,superscriptaddress
]{revtex4-2}

\usepackage{graphicx}
\usepackage{dcolumn}
\usepackage{bm}
\usepackage{url}
\usepackage{float}

\begin{document}

\preprint{}

\title{Search for a Non-Relativistic Boson in Two-Body Antimuon Decay}

\author{J.I.\ Collar}
\email{collar@uchicago.edu}
\affiliation{Enrico Fermi Institute, Kavli Institute for Cosmological Physics, and Department of Physics\\
University of Chicago, Chicago, Illinois 60637, USA}

\affiliation{Donostia International Physics Center (DIPC), Paseo Manuel Lardizabal 4, 20018 Donostia-San Sebastian, Spain}

\author{P.S.\ Cooper}
\email{pcooper.fnal@gmail.com}
\affiliation{Fermi National Accelerator Laboratory, Batavia, Illinois 60510, USA}

\author{C.M.\ Lewis}
\email{mark.lewis@dipc.org}
\affiliation{Enrico Fermi Institute, Kavli Institute for Cosmological Physics, and Department of Physics\\
University of Chicago, Chicago, Illinois 60637, USA}

\affiliation{Donostia International Physics Center (DIPC), Paseo Manuel Lardizabal 4, 20018 Donostia-San Sebastian, Spain}



\date{\today}

\begin{abstract}
We demonstrate the feasibility of probing the charged lepton flavor violating decay $\mu^{+}\!\!\rightarrow \!e^{+} X^{0}$ for the presence of a slow-moving  neutral boson $X^{0}$ capable of undergoing gravitational binding to large structures, and as such able to participate in some cosmological scenarios. A short exposure to surface antimuons from beamline M20 at TRIUMF generates a branching ratio limit  of $\lesssim 10^{-5}$. This is comparable or better than  previous searches for this channel, although in a thus-far unexplored region of $X^{0}$ phase space  very close to the kinematic limit of the decay. The future improved   sensitivity of the method using a customized p-type point contact germanium detector is described. 
\end{abstract}

\maketitle


The strongest present evidence for the incompleteness of the Standard Model (SM) arises from the observation of lepton-flavor violation in the neutrino sector, manifested through the phenomenon of  neutrino oscillations. This neutral-particle precedent guarantees the eventual appearance of charged lepton flavor violation (CLFV) \cite{clfv,prospects2}, its detection  within reach according to some favorable phenomenological perspectives \cite{prospects1,prospects2}.  

Numerous extensions of the SM generate new massive neutral bosons $X^{0}$ with lepton-flavor-violating couplings, e.g., axion or axion-like particles \cite{axion1,axion2,axion3}, Majorons and Familons \cite{majoron1,majoron2,majoron3,majoron4}, light gauge bosons Z$'$ \cite{Z1,Z2,Z3}, etc. These have been sought in kaon \cite{kaon1,kaon2,kaon3,kaon4,kaon5,kaon6}, pion \cite{pion1,pion2,pion3}, and tau \cite{tau1,tau2} decays. Muon decay is unique in that it proceeds exclusively via the weak interaction, resulting in a well-understood, single known mode ($\mu\rightarrow e \bar{\nu_{e}} \nu_{\mu}$, with radiative derivatives such as $\mu\rightarrow e \bar{\nu_{e}} \nu_{\mu} \gamma$). In this way, it offers what is arguably the simplest framework to look for  deviations involving CLFV.  Not surprisingly, multiple searches for $\mu^{+}\!\!\rightarrow \!e^{+} X^{0}$ have been carried out \cite{search1,search2,search3,search4,search5,search6,search7,search8,pion1}.

In certain models \cite{mot3} the two-body decay $\mu\!\!\rightarrow \!e X^{0}$ is favored over other CLFV alternatives such as $\mu\rightarrow e \gamma$, $\mu\rightarrow 3e$ or  $\mu N\rightarrow e N$ \cite{clfv1}. Striking evidence for $\mu^{+}\!\!\rightarrow \!e^{+} X^{0}$ would be provided by an anomalous  peak embedded in the positron continuum (Michel spectrum \cite{michel1}) from the dominating $\mu^{+}\rightarrow e^{+} \nu_{e} \bar{\nu_{\mu}}$. Sensitivity to this signature has nevertheless been affected by the moderate energy resolution of the calorimeters employed in  previous searches, typically  inorganic scintillators. It has been recently proposed \cite{myprd} to use p-type point-contact (PPC) germanium detector \cite{ppc} technology to bypass this shortcoming while simultaneously entering the regime of  boson emission speed that can lead to its gravitational binding to astronomical objects, in the guise of a dark matter. Of  mention is the possibility that one such $X^{0}$, sufficiently long-lived and  with rest mass close to that of the antimuon could generate through its decay ($X^{0} \rightarrow e^{+} e^{-} \bar{\nu} \nu$, $X^{0} \rightarrow e^{+} e^{-} \phi$ \cite{sergei}) the positron injection energy  \cite{john,john2,5113} characteristic of the 511 keV gamma emission from the bulge of the Milky Way \cite{myprd}. Its origin is an unresolved mystery spanning five decades \cite{5111,5112}. 

\begin{figure}[!htbp]
\includegraphics[width=.75 \linewidth]{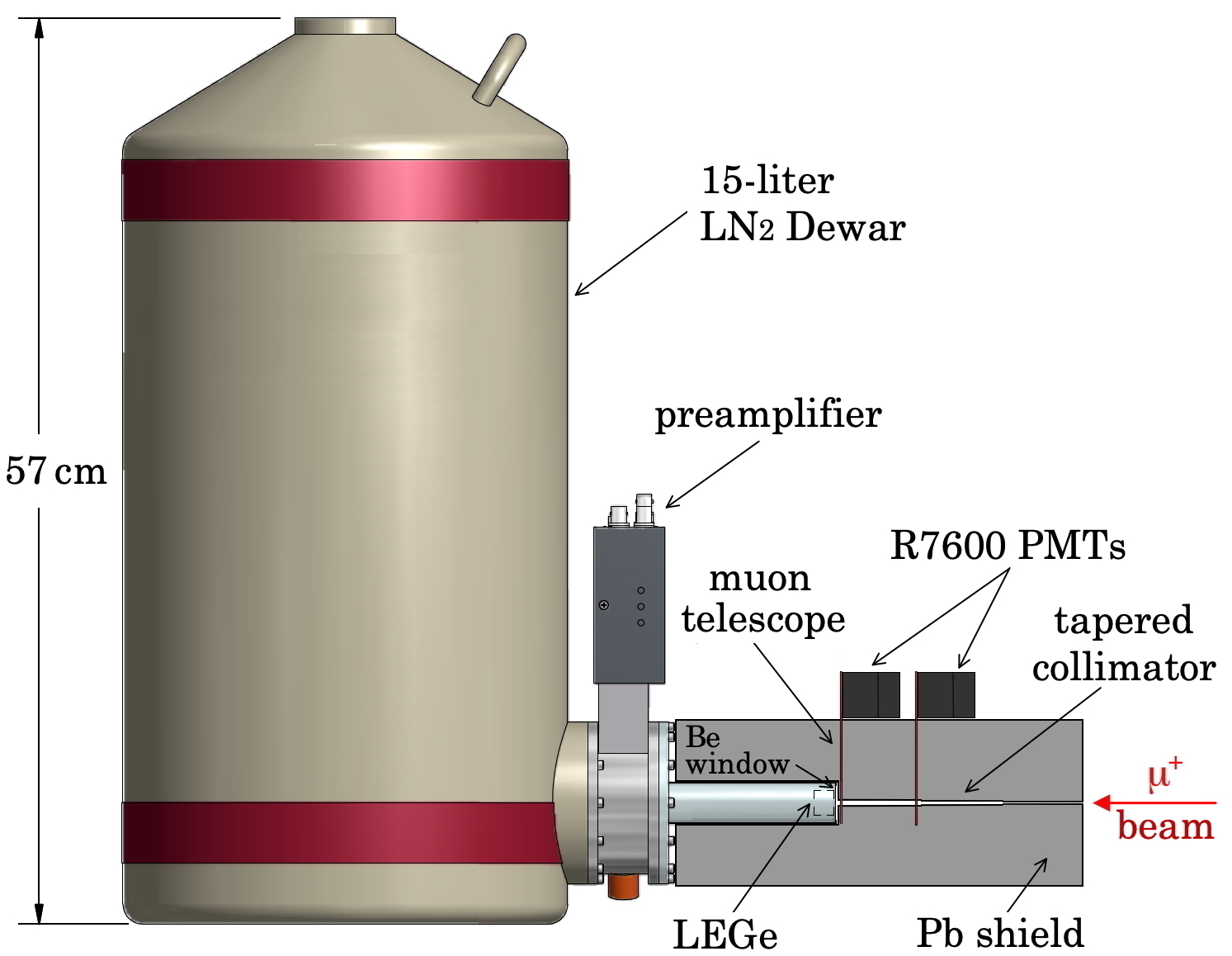}
\caption{\label{fig:epsart} Schematic of the apparatus. Surface $\mu^{+}$ are stopped in a 0.25 cm$^{3}$ n-type germanium detector (LEGe \cite{lege}). A  trigger is provided by an ultra-thin muon telescope. The device fits on a 60 cm $\times$ 30 cm tray, weights 
less than 50 kg, and is therefore somewhat too small to be 
described as ``table top".}
\end{figure}

Ref.\ \cite{myprd} provides more information on the motivation,  implementation, and sensitivity of this search. Briefly stated, the present approach involves the stopping of low energy (\mbox{$E_{\mu^{+}}\approx$ 4.1 MeV}) ``surface" antimuons \cite{surface} in an active  ``germanium beam dump" (Fig.\ 1), leading to the detection of $E_{\mu^{+}}$ and positron kinetic energy $E_{e^{+}}$ in rapid succession ($\tau_{\mu}\!\!=\!2.197 \mu$s). This is challenging, specially in the $E_{e^{+}}\!<\!<\!E_{\mu^{+}}$ domain where the $X^{0}$ is non-relativistic. As discussed below, high-throughput signal digitization followed by a sophisticated offline analysis has allowed for a first exploration in this type of search of the few-keV $E_{e^{+}}$ range. This corresponds to a $X^{0}$ travelling slower than escape velocities from stars and galaxies \cite{myprd}.   

The modest energy characteristic of surface muons leads to their shallow implantation in the front surface of the germanium detector, where their decay at rest proceeds. As shown below, in order to minimize backgrounds in the $E_{e^{+}}$ range of interest, it is imperative to reduce any in-flight  degradation of $E_{\mu^{+}}$  \cite{myprd}. This was accomplished by using a thin (25 $\mu$m) Kapton exit window on TRIUMF's M20 beamline \cite{triumf}. The assembly in Fig.\ 1 was placed immediately next to it, using alignment lasers and an adjustable platform to ensure the coaxiality of collimator and beam. The telescope responsible for providing a data acquisition (DAQ) trigger on muon crossing introduced just two layers of 16 $\mu$m Al foil and a 25 $\mu$m film of plastic scintillator \cite{eljen} into the  beam path. The high light-detection efficiency of ultra-bialkali R7600U-200 photomultipliers (PMTs) permitted the use of a single telescope paddle operated slightly above single photoelectron sensitivity. A 25 $\mu$m Be window at the entrance of the detector cryostat and 17 cm of air were the only other materials penetrated by the collimated beam. This led to a tolerable 20\% muon energy loss (Fig.\ 2).

\begin{figure}[!htbp]
\includegraphics[width=.77 \linewidth]{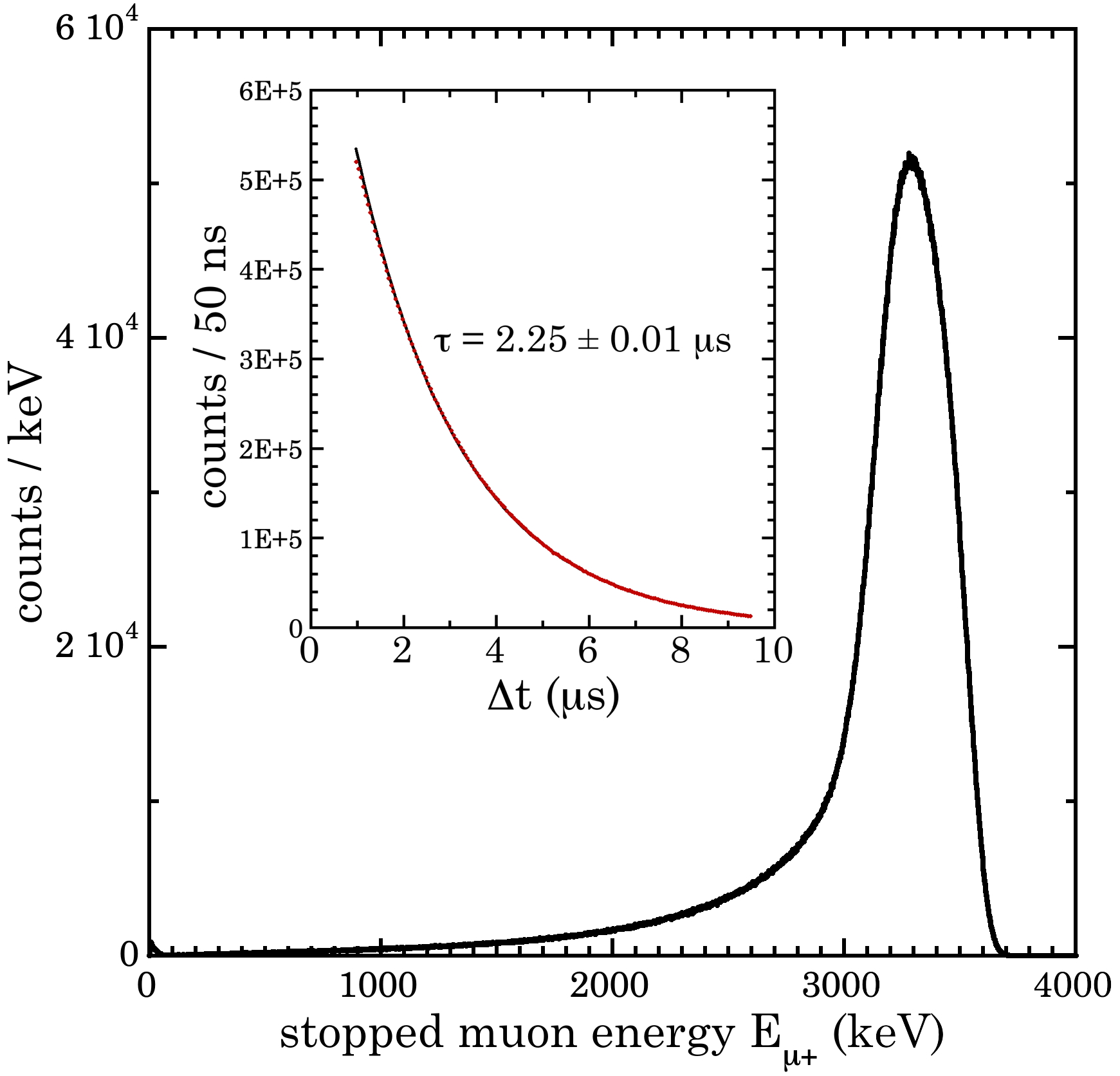}
\caption{\label{fig:epsart} $\mu^{+}$ energy measured by the LEGe detector,  degraded from the nominal beam value (4.1 MeV) due to intermediary material traversal. {\it Inset:}  distribution of intervals $\Delta$t between $\mu^{+}$ and e$^{+}$ signals. The small variance between the best-fit value shown and $\tau_{\mu}=2.197 ~\mu$s is traceable to a known systematic affecting e$^{+}$ and $\mu^{+}$ pulse onset determination differently. }
\end{figure}

To avoid multiple difficulties derived from reliance on analog electronics in a previous exposure of a germanium detector to an antimuon beam \cite{search3}, LEGe raw preamplifier traces were saved to disk for each event. Beam focusing and Pb collimator aperture were optimized to provide a  trigger rate of up to 1,800 Hz, devoid of pile-up, while also within the throughput capabilities of a GaGe RazorMax 161G40 digitizer. In this initial run, the intrinsic noise of the available digitizer (larger than that of the LEGe detector by a factor of $\sim20$) limited $E_{e^{+}}$ sensitivity to energies above 3 keV. This also impacted the best energy resolution achievable (Fig.\ 4). 

Fig.\ 3 displays example event traces. A saved \mbox{10 $\mu$s} pre-trigger segment allowed to study environmental and beam-induced backgrounds. These were negligible \cite{mark}, demonstrating the adequacy of  shielding design \cite{myprd}. This ability to inspect anti-coincident backgrounds would nevertheless be useful to clarify the origin of any anomalous peaks found in the $E_{e^{+}}$ spectrum \cite{myprd}.

\begin{figure}[!htbp]
\includegraphics[width=.79 \linewidth]{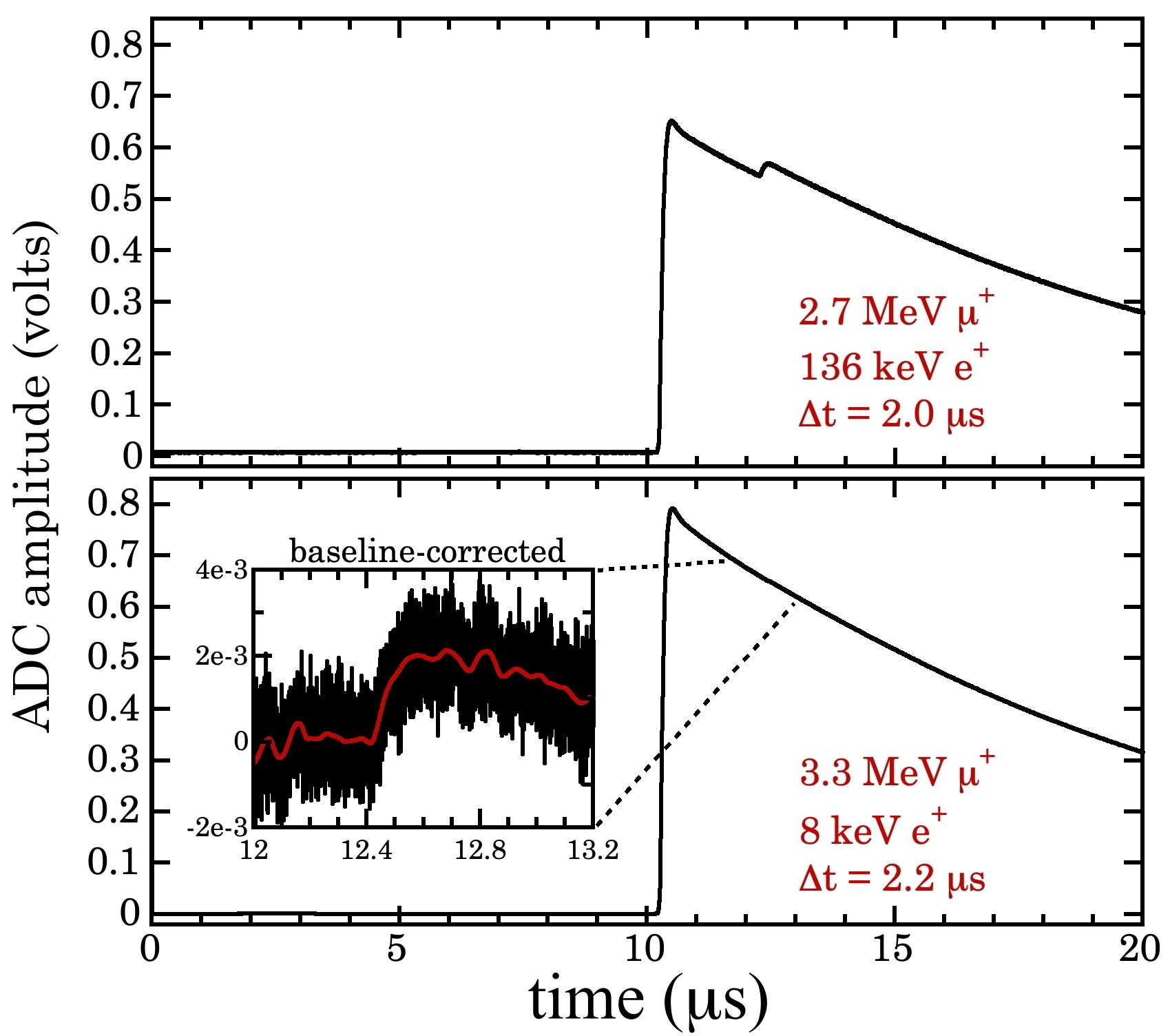}
\caption{\label{fig:epsart} Two example events. The one in the bottom panel is close to the limit of $E_{e^{+}}$ detectability for this run. A red line in the inset shows the denoised trace around the moment of  e$^{+}$ emission. The value of an in-line capacitor, responsible for the  pulse decay visible, was adjusted to provide a compromise between  $E_{e^{+}}$ threshold and preamplifier reset rate \cite{myprd}.}
\end{figure}

Techniques developed for the identification of sub-keV germanium detector signals were applied during off-line analysis, with emphasis on extending the positron spectrum to the lowest possible energies. These include wavelet denoising, optimized digital shaping filters for energy measurement, and an edge-finding algorithm for signal onset determination. They have been recently described in the context of a  first measurement of coherent elastic neutrino-nucleus scattering from reactor antineutrinos \cite{dresden1,dresden2}. Their use allowed to identify positron signals as small as few keV in close temporal proximity to the much larger muon-stopping pulse. Specifically, within the span \mbox{0.95 $\mu$s $< \Delta$t $<$ 9.5 $\mu$s} (Fig.\ 2, inset), its boundaries defined by excessive overlap of e$^{+}$ and $\mu^{+}$ shaped signals and by digitized trace length, respectively. This provides a satisfactory 64\%  muon decay acceptance. The LEGe energy scale and resolution were studied using gamma emissions from $^{133}$Ba in a calibration performed immediately before  M20 data-taking \cite{mark}.

Beam exposure was limited to two days with a slow ramp up in power reaching a maximum of just 60\% of the nominal muon current available at M20, the result of beam magnet problems during the start of schedule 142 at TRIUMF. Nevertheless, the high data throughput achieved allowed to gather a total of $1.3\times10^{8}$ triggers, of which $2.0\times10^{7}$ passed quality (spurious triggers from PMT dark current,  trace contamination by preamplifier resets) and $\Delta$t cuts. 

Fig.\ 4 shows $E_{e^{+}}$ spectra for several choices of minimum $E_{\mu^{+}}$ accepted. A rapid rise in background can be observed to migrate towards higher $E_{e^{+}}$ as this selection varies. It originates in positrons emitted in the opposite direction to incoming muons, able to escape the germanium crystal before they deposit their full energy \cite{myprd}. Their energy loss correlates to muon stopping depth and in turn to $E_{\mu^{+}}$. A  compromise between background reduction in the $E_{e^{+}}$ region of interest and  signal statistics available, leading to the best possible $\mu^{+}\!\!\rightarrow \!e^{+} X^{0}$ branching ratio (BR) sensitivity, was found for \mbox{$E_{\mu^{+}} > 3.35$ MeV}. This provides a 35\% event acceptance (Fig.\ 2).

\begin{figure}[!htbp]
\includegraphics[width=.85 \linewidth]{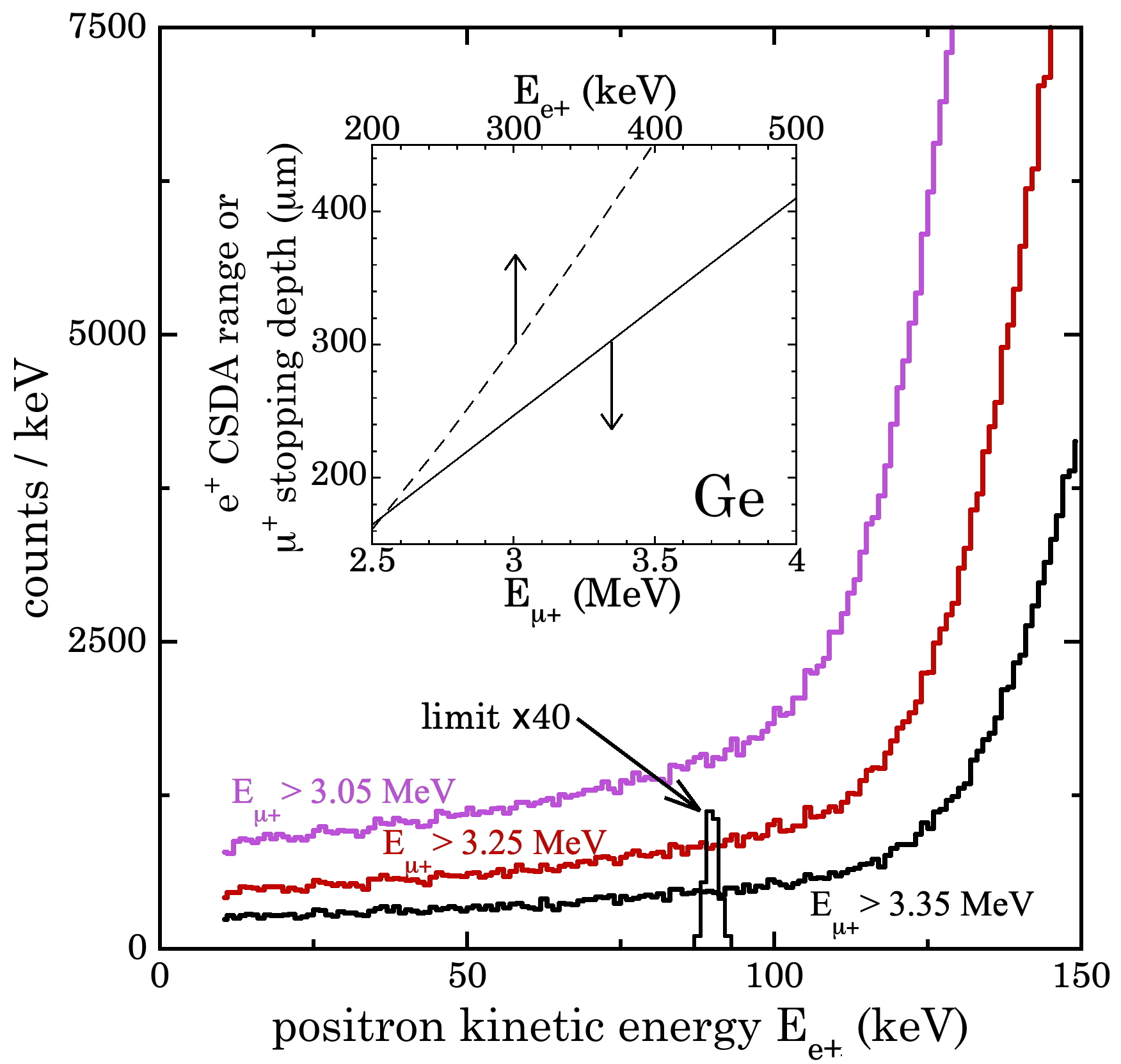}
\caption{\label{fig:epsart} Positron energy spectra for labelled choices of antimuon energy acceptance. A rising background is due to partial energy depositions from  e$^{+}$ germanium escape \cite{myprd}. The energy resolution achieved is illustrated by an overlapped peak with intensity corresponding to $\times$40 the present sensitivity of the search. See text for a discussion on the inset. }
\end{figure}

The inset of Fig.\ 4 shows the continuous-slow-down approximation (CSDA) e$^{+}$ range \cite{csda1,csda2} and MCNPX \cite{mcnpx} simulated average depth of $\mu^{+}$ implantation, both for germanium and as a function of initial particle energy. Naively, it might be concluded from it that only positrons with \mbox{$E_{e^{+}} \lesssim 300$ keV} can be guaranteed to lose all their energy within the germanium crystal as $E_{\mu^{+}}$ approaches 3.35 MeV. However,  positrons show simulated trajectories fully contained in germanium for \mbox{$E_{e^{+}} < 400$ keV} when originated at the $\sim$300 $\mu$m  implantation depth characteristic of \mbox{$E_{\mu^{+}} = 3.35$ MeV} \cite{mark}. This is the result of a positron detour factor (ratio of projected range to  CSDA range) significantly smaller than unity in this energy range. For simplicity and in view of the much improved future expected sensitivity discussed below, we restrict the present analysis to fully-contained e$^{+}$ tracks  (\mbox{$E_{e^{+}} < 400$ keV} for \mbox{$E_{\mu^{+}} > 3.35$ MeV}). The $\sim$10\% probability of interaction for e$^{+}$ annihilation  radiation (511 keV) in this small detector  is also neglected \cite{myprd}. Visual inspection of the positron spectrum out 5 MeV, close to the simulated maximum energy loss  in a germanium crystal of this size, finds no obvious peak-like features. 

An unbinned search algorithm \cite{mark} reveals no significant peak-like structures in the examined region of the positron spectrum (Fig.\ 5, top panel). Energy resolution and local background vary considerably over the \mbox{ 9 keV $< E_{e^{+}} < 400$ keV} range analyzed. Specifically, the full-width-at-half-maximum (FWHM) resolution grows from 2.1 keV to 4.8 keV over this interval. Both factors are included in the algorithm, which generates the statistical significance of any peak-like accumulation of events via Monte Carlo calculation. The present 95\% C.L. BR sensitivity of the search (Fig.\ 5, bottom panel) is estimated as $2\sqrt{N}/(0.76\cdot T)$, where $N$ is the number of positron counts at any energy within a window corresponding to the local FWHM resolution (containing 76\% of counts under a peak) and  $T=7.1\times10^{6}$ is total events passing all cuts, including that imposed on $E_{\mu^{+}}$.  

\begin{figure}[!htbp]
\includegraphics[width=1. \linewidth]{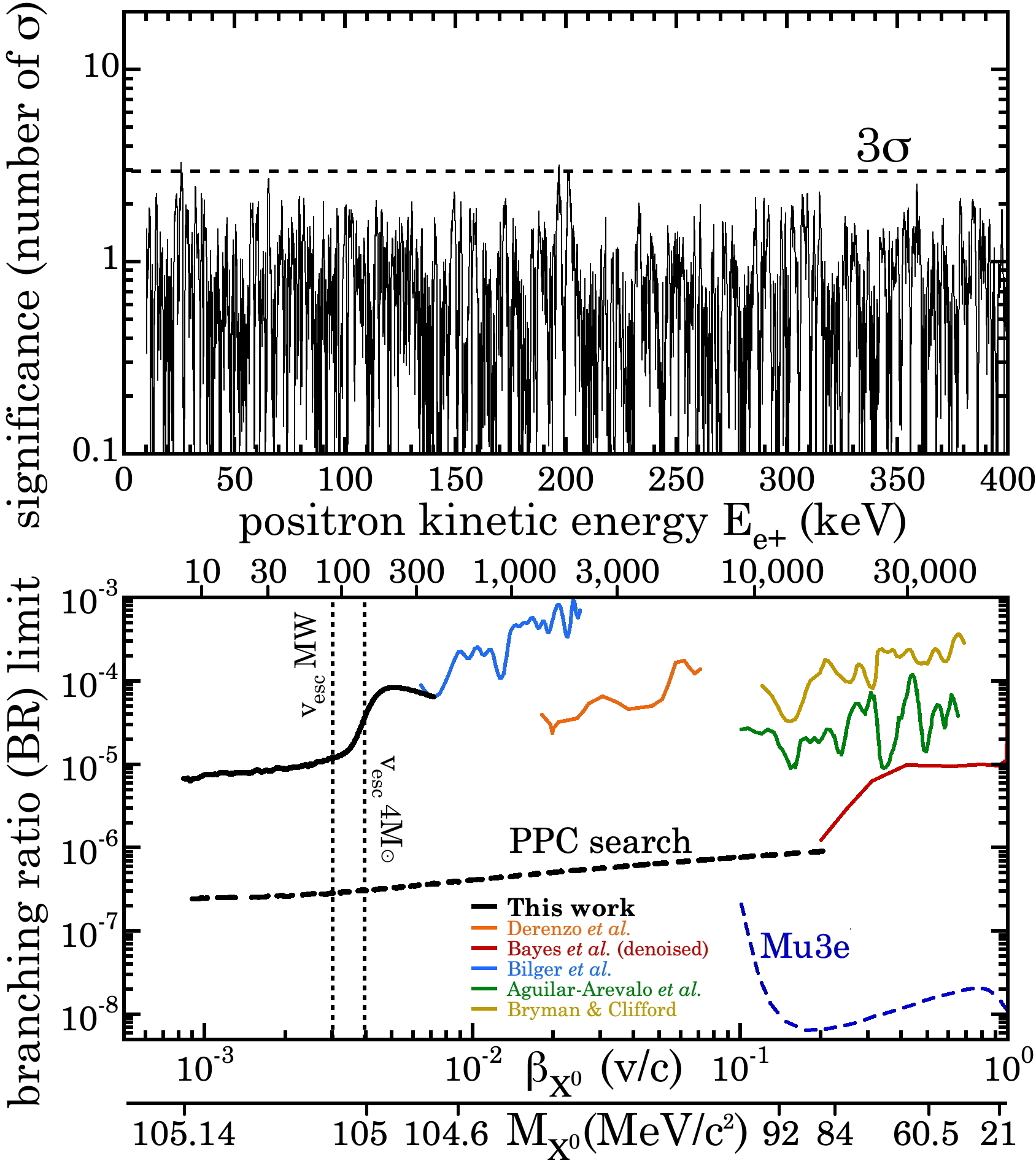}
\caption{\label{fig:epsart} {\it Top:} negative result from an unbinned peak search in the e$^{+}$ energy region of interest, consistent with statistical fluctuations. {\it Bottom:} previous and present BR upper limits (solid lines) for $\mu^{+}\!\!\rightarrow \!e^{+} X^{0}$ versus speed of the emitted boson $\beta_{X^{0}}$ (its rest mass M$_{X^{0}}$ is also indicated). Vertical lines indicate escape velocities from the gravitational well of the Milky Way (MW) at its center and from a massive star. Dashed lines represent expectations for this method, under  conditions described in the text, and for Mu3e \cite{mu3e}.  }
\end{figure}

Fig.\ 5 also displays the expected future sensitivity of the ``germanium beam dump" method in a final run using higher energy ($\sim$37 MeV) ``cloud" muons stopped in a  $\sim$50 cm$^{3}$ PPC, presently under development. A p-type contact configuration is necessary to avoid the grave degradation of resolution  characteristic of large n-type point contact diodes \cite{Luke,myprd}. The dominating low-energy background in this configuration, where muons are deeply implanted into the core of the device, is expected to arise from infrequent positron annihilation in flight \cite{myprd}. The expectation shown is calculated under this premise, assuming a five-day beam exposure and a conservative trigger rate of 1,000 Hz. The effects of a reduced digitizer noise and a decreased preamplifier gain (necessary to accommodate broader e$^{+}$ and $\mu^{+}$ energy ranges) are expected to offset each other, leading to a similar few-keV $E_{e^{+}}$ threshold. For positron energies above $\sim$20 MeV, radiative losses and maximum PPC size are expected to severely diminish the sensitivity of this method \cite{myprd}. Coaxial  geometries leading to larger (up to $\sim\times$10) point-contact crystals \cite{dresden1} are not considered due to their impact on the containment of positron tracks. 

The emergence of detector technologies with sensitivity to lower energies and improved resolution  invites a shift of emphasis in particle-decay searches for massive neutral bosons: the kinematic limit of these reactions remains an unexplored realm where cosmologically-relevant particles may lie in wait.

We thank TRIUMF and its personnel for  making this work possible. Specifically, Rahim Abasalti, Friedhelm Ames, Donald Arseneau, Martin Alcorta Moreno, Jens Dilling, Sarah Dunsiger, Bassam Hitti, Angela Lang, Iain McKenzie, Gerald Morris, Nigel Smith, Oliver Stelzer-Chilton, Deepak Vyas, and many others. Our gratitude also goes to Rocky Kolb for providing much needed initial support for this effort and to Doug Bryman, Jim Colaresi, and Wilhelm Mueller  for useful exchanges. This work is funded by NSF award PHY-2209456 and the
Kavli Institute for Cosmological Physics at the University of Chicago, through an endowment from the Kavli Foundation and its founder Fred Kavli.

\bibliography{apssamp}

\providecommand{\noopsort}[1]{}\providecommand{\singleletter}[1]{#1}%
\begin{thebibliography}{55}%
\makeatletter
\providecommand \@ifxundefined [1]{%
 \@ifx{#1\undefined}
}%
\providecommand \@ifnum [1]{%
 \ifnum #1\expandafter \@firstoftwo
 \else \expandafter \@secondoftwo
 \fi
}%
\providecommand \@ifx [1]{%
 \ifx #1\expandafter \@firstoftwo
 \else \expandafter \@secondoftwo
 \fi
}%
\providecommand \natexlab [1]{#1}%
\providecommand \enquote  [1]{``#1''}%
\providecommand \bibnamefont  [1]{#1}%
\providecommand \bibfnamefont [1]{#1}%
\providecommand \citenamefont [1]{#1}%
\providecommand \href@noop [0]{\@secondoftwo}%
\providecommand \href [0]{\begingroup \@sanitize@url \@href}%
\providecommand \@href[1]{\@@startlink{#1}\@@href}%
\providecommand \@@href[1]{\endgroup#1\@@endlink}%
\providecommand \@sanitize@url [0]{\catcode `\\12\catcode `\$12\catcode
  `\&12\catcode `\#12\catcode `\^12\catcode `\_12\catcode `\%12\relax}%
\providecommand \@@startlink[1]{}%
\providecommand \@@endlink[0]{}%
\providecommand \url  [0]{\begingroup\@sanitize@url \@url }%
\providecommand \@url [1]{\endgroup\@href {#1}{\urlprefix }}%
\providecommand \urlprefix  [0]{URL }%
\providecommand \Eprint [0]{\href }%
\providecommand \doibase [0]{https://doi.org/}%
\providecommand \selectlanguage [0]{\@gobble}%
\providecommand \bibinfo  [0]{\@secondoftwo}%
\providecommand \bibfield  [0]{\@secondoftwo}%
\providecommand \translation [1]{[#1]}%
\providecommand \BibitemOpen [0]{}%
\providecommand \bibitemStop [0]{}%
\providecommand \bibitemNoStop [0]{.\EOS\space}%
\providecommand \EOS [0]{\spacefactor3000\relax}%
\providecommand \BibitemShut  [1]{\csname bibitem#1\endcsname}%
\let\auto@bib@innerbib\@empty
\bibitem [{\citenamefont {Davidson}\ \emph {et~al.}()\citenamefont {Davidson},
  \citenamefont {Echenard}, \citenamefont {Bernstein}, \citenamefont {Heeck},\
  and\ \citenamefont {Hitlin}}]{clfv}%
  \BibitemOpen
  \bibfield  {author} {\bibinfo {author} {\bibfnamefont {S.}~\bibnamefont
  {Davidson}}, \bibinfo {author} {\bibfnamefont {B.}~\bibnamefont {Echenard}},
  \bibinfo {author} {\bibfnamefont {R.~H.}\ \bibnamefont {Bernstein}}, \bibinfo
  {author} {\bibfnamefont {J.}~\bibnamefont {Heeck}},\ and\ \bibinfo {author}
  {\bibfnamefont {D.~G.}\ \bibnamefont {Hitlin}},\ }\href@noop {} {}\Eprint
  {https://arxiv.org/abs/arXiv:2209.00142} {arXiv:2209.00142} \BibitemShut
  {NoStop}%
\bibitem [{\citenamefont {Hambye}(2014)}]{prospects2}%
  \BibitemOpen
  \bibfield  {author} {\bibinfo {author} {\bibfnamefont {T.}~\bibnamefont
  {Hambye}},\ }\href@noop {} {\bibfield  {journal} {\bibinfo  {journal} {Nucl.
  Phys. B Proc. Suppl.}\ }\textbf {\bibinfo {volume} {248-250}},\ \bibinfo
  {pages} {13 } (\bibinfo {year} {2014})}\BibitemShut {NoStop}%
\bibitem [{\citenamefont {Hirsch}\ \emph {et~al.}(2009)\citenamefont {Hirsch},
  \citenamefont {Vicente}, \citenamefont {Meyer},\ and\ \citenamefont
  {Porod}}]{prospects1}%
  \BibitemOpen
  \bibfield  {author} {\bibinfo {author} {\bibfnamefont {M.}~\bibnamefont
  {Hirsch}}, \bibinfo {author} {\bibfnamefont {A.}~\bibnamefont {Vicente}},
  \bibinfo {author} {\bibfnamefont {J.}~\bibnamefont {Meyer}},\ and\ \bibinfo
  {author} {\bibfnamefont {W.}~\bibnamefont {Porod}},\ }\href@noop {}
  {\bibfield  {journal} {\bibinfo  {journal} {Phys. Rev. D}\ }\textbf {\bibinfo
  {volume} {79}},\ \bibinfo {pages} {055023} (\bibinfo {year}
  {2009})}\BibitemShut {NoStop}%
\bibitem [{\citenamefont {Jaeckel}\ and\ \citenamefont
  {Ringwald}(2010)}]{axion1}%
  \BibitemOpen
  \bibfield  {author} {\bibinfo {author} {\bibfnamefont {J.}~\bibnamefont
  {Jaeckel}}\ and\ \bibinfo {author} {\bibfnamefont {A.}~\bibnamefont
  {Ringwald}},\ }\href@noop {} {\bibfield  {journal} {\bibinfo  {journal}
  {Annu. Rev. Nucl. Part.}\ }\textbf {\bibinfo {volume} {60}},\ \bibinfo
  {pages} {405} (\bibinfo {year} {2010})}\BibitemShut {NoStop}%
\bibitem [{\citenamefont {Choi}\ \emph {et~al.}(2021)\citenamefont {Choi},
  \citenamefont {Im},\ and\ \citenamefont {Shin}}]{axion2}%
  \BibitemOpen
  \bibfield  {author} {\bibinfo {author} {\bibfnamefont {K.}~\bibnamefont
  {Choi}}, \bibinfo {author} {\bibfnamefont {S.~H.}\ \bibnamefont {Im}},\ and\
  \bibinfo {author} {\bibfnamefont {C.~S.}\ \bibnamefont {Shin}},\ }\href@noop
  {} {\bibfield  {journal} {\bibinfo  {journal} {Annu. Rev. Nucl. Part.}\
  }\textbf {\bibinfo {volume} {71}},\ \bibinfo {pages} {225} (\bibinfo {year}
  {2021})}\BibitemShut {NoStop}%
\bibitem [{\citenamefont {Alves}\ and\ \citenamefont {Weiner}(2018)}]{axion3}%
  \BibitemOpen
  \bibfield  {author} {\bibinfo {author} {\bibfnamefont {D.~S.~M.}\
  \bibnamefont {Alves}}\ and\ \bibinfo {author} {\bibfnamefont
  {N.}~\bibnamefont {Weiner}},\ }\href@noop {} {\bibfield  {journal} {\bibinfo
  {journal} {Journal of High Energy Physics}\ }\textbf {\bibinfo {volume}
  {07}},\ \bibinfo {pages} {092} (\bibinfo {year} {2018})}\BibitemShut
  {NoStop}%
\bibitem [{\citenamefont {Gelmini}\ and\ \citenamefont
  {Roncadelli}(1981)}]{majoron1}%
  \BibitemOpen
  \bibfield  {author} {\bibinfo {author} {\bibfnamefont {G.}~\bibnamefont
  {Gelmini}}\ and\ \bibinfo {author} {\bibfnamefont {M.}~\bibnamefont
  {Roncadelli}},\ }\href@noop {} {\bibfield  {journal} {\bibinfo  {journal}
  {Phys. Lett. B}\ }\textbf {\bibinfo {volume} {99}},\ \bibinfo {pages} {411}
  (\bibinfo {year} {1981})}\BibitemShut {NoStop}%
\bibitem [{\citenamefont {Chikashige}\ \emph {et~al.}(1981)\citenamefont
  {Chikashige}, \citenamefont {Mohapatra},\ and\ \citenamefont
  {Peccei}}]{majoron2}%
  \BibitemOpen
  \bibfield  {author} {\bibinfo {author} {\bibfnamefont {Y.}~\bibnamefont
  {Chikashige}}, \bibinfo {author} {\bibfnamefont {R.}~\bibnamefont
  {Mohapatra}},\ and\ \bibinfo {author} {\bibfnamefont {R.}~\bibnamefont
  {Peccei}},\ }\href@noop {} {\bibfield  {journal} {\bibinfo  {journal} {Phys.
  Lett. B}\ }\textbf {\bibinfo {volume} {98}},\ \bibinfo {pages} {265}
  (\bibinfo {year} {1981})}\BibitemShut {NoStop}%
\bibitem [{\citenamefont {Aulakh}\ and\ \citenamefont
  {Mohapatra}(1982)}]{majoron3}%
  \BibitemOpen
  \bibfield  {author} {\bibinfo {author} {\bibfnamefont {C.}~\bibnamefont
  {Aulakh}}\ and\ \bibinfo {author} {\bibfnamefont {R.}~\bibnamefont
  {Mohapatra}},\ }\href@noop {} {\bibfield  {journal} {\bibinfo  {journal}
  {Phys. Lett. B}\ }\textbf {\bibinfo {volume} {119}},\ \bibinfo {pages} {136}
  (\bibinfo {year} {1982})}\BibitemShut {NoStop}%
\bibitem [{\citenamefont {Wilczek}(1982)}]{majoron4}%
  \BibitemOpen
  \bibfield  {author} {\bibinfo {author} {\bibfnamefont {F.}~\bibnamefont
  {Wilczek}},\ }\href@noop {} {\bibfield  {journal} {\bibinfo  {journal} {Phys.
  Rev. Lett.}\ }\textbf {\bibinfo {volume} {49}},\ \bibinfo {pages} {1549}
  (\bibinfo {year} {1982})}\BibitemShut {NoStop}%
\bibitem [{\citenamefont {Foot}\ \emph {et~al.}(1994)\citenamefont {Foot},
  \citenamefont {He}, \citenamefont {Lew},\ and\ \citenamefont {Volkas}}]{Z1}%
  \BibitemOpen
  \bibfield  {author} {\bibinfo {author} {\bibfnamefont {R.}~\bibnamefont
  {Foot}}, \bibinfo {author} {\bibfnamefont {X.-G.}\ \bibnamefont {He}},
  \bibinfo {author} {\bibfnamefont {H.}~\bibnamefont {Lew}},\ and\ \bibinfo
  {author} {\bibfnamefont {R.~R.}\ \bibnamefont {Volkas}},\ }\href@noop {}
  {\bibfield  {journal} {\bibinfo  {journal} {Phys. Rev. D}\ }\textbf {\bibinfo
  {volume} {50}},\ \bibinfo {pages} {4571} (\bibinfo {year}
  {1994})}\BibitemShut {NoStop}%
\bibitem [{\citenamefont {Heeck}(2016)}]{Z2}%
  \BibitemOpen
  \bibfield  {author} {\bibinfo {author} {\bibfnamefont {J.}~\bibnamefont
  {Heeck}},\ }\href@noop {} {\bibfield  {journal} {\bibinfo  {journal} {Phys.
  Lett. B}\ }\textbf {\bibinfo {volume} {758}},\ \bibinfo {pages} {101}
  (\bibinfo {year} {2016})}\BibitemShut {NoStop}%
\bibitem [{\citenamefont {Altmannshofer}\ \emph {et~al.}(2016)\citenamefont
  {Altmannshofer}, \citenamefont {Chen}, \citenamefont {Dev},\ and\
  \citenamefont {Soni}}]{Z3}%
  \BibitemOpen
  \bibfield  {author} {\bibinfo {author} {\bibfnamefont {W.}~\bibnamefont
  {Altmannshofer}}, \bibinfo {author} {\bibfnamefont {C.-Y.}\ \bibnamefont
  {Chen}}, \bibinfo {author} {\bibfnamefont {P.~B.}\ \bibnamefont {Dev}},\ and\
  \bibinfo {author} {\bibfnamefont {A.}~\bibnamefont {Soni}},\ }\href@noop {}
  {\bibfield  {journal} {\bibinfo  {journal} {Phys. Lett. B}\ }\textbf
  {\bibinfo {volume} {762}},\ \bibinfo {pages} {389} (\bibinfo {year}
  {2016})}\BibitemShut {NoStop}%
\bibitem [{\citenamefont {Yamazaki}\ \emph {et~al.}(1984)\citenamefont
  {Yamazaki} \emph {et~al.}}]{kaon1}%
  \BibitemOpen
  \bibfield  {author} {\bibinfo {author} {\bibfnamefont {T.}~\bibnamefont
  {Yamazaki}} \emph {et~al.},\ }\href@noop {} {\bibfield  {journal} {\bibinfo
  {journal} {Phys. Rev. Lett.}\ }\textbf {\bibinfo {volume} {52}},\ \bibinfo
  {pages} {1089} (\bibinfo {year} {1984})}\BibitemShut {NoStop}%
\bibitem [{\citenamefont {Baker}\ \emph {et~al.}(1987)\citenamefont {Baker}
  \emph {et~al.}}]{kaon2}%
  \BibitemOpen
  \bibfield  {author} {\bibinfo {author} {\bibfnamefont {N.~J.}\ \bibnamefont
  {Baker}} \emph {et~al.},\ }\href@noop {} {\bibfield  {journal} {\bibinfo
  {journal} {Phys. Rev. Lett.}\ }\textbf {\bibinfo {volume} {59}},\ \bibinfo
  {pages} {2832} (\bibinfo {year} {1987})}\BibitemShut {NoStop}%
\bibitem [{\citenamefont {Adler}\ \emph {et~al.}(2002)\citenamefont {Adler}
  \emph {et~al.}}]{kaon3}%
  \BibitemOpen
  \bibfield  {author} {\bibinfo {author} {\bibfnamefont {S.}~\bibnamefont
  {Adler}} \emph {et~al.},\ }\href@noop {} {\bibfield  {journal} {\bibinfo
  {journal} {Phys. Lett. B}\ }\textbf {\bibinfo {volume} {537}},\ \bibinfo
  {pages} {211} (\bibinfo {year} {2002})}\BibitemShut {NoStop}%
\bibitem [{\citenamefont {Adler}\ \emph {et~al.}(2004)\citenamefont {Adler}
  \emph {et~al.}}]{kaon4}%
  \BibitemOpen
  \bibfield  {author} {\bibinfo {author} {\bibfnamefont {S.}~\bibnamefont
  {Adler}} \emph {et~al.},\ }\href@noop {} {\bibfield  {journal} {\bibinfo
  {journal} {Phys. Rev. D}\ }\textbf {\bibinfo {volume} {70}},\ \bibinfo
  {pages} {037102} (\bibinfo {year} {2004})}\BibitemShut {NoStop}%
\bibitem [{\citenamefont {Anisimovsky}\ \emph {et~al.}(2004)\citenamefont
  {Anisimovsky} \emph {et~al.}}]{kaon5}%
  \BibitemOpen
  \bibfield  {author} {\bibinfo {author} {\bibfnamefont {V.~V.}\ \bibnamefont
  {Anisimovsky}} \emph {et~al.},\ }\href@noop {} {\bibfield  {journal}
  {\bibinfo  {journal} {Phys. Rev. Lett.}\ }\textbf {\bibinfo {volume} {93}},\
  \bibinfo {pages} {031801} (\bibinfo {year} {2004})}\BibitemShut {NoStop}%
\bibitem [{\citenamefont {Atiya}\ \emph {et~al.}(1990)\citenamefont {Atiya}
  \emph {et~al.}}]{kaon6}%
  \BibitemOpen
  \bibfield  {author} {\bibinfo {author} {\bibfnamefont {M.~S.}\ \bibnamefont
  {Atiya}} \emph {et~al.},\ }\href@noop {} {\bibfield  {journal} {\bibinfo
  {journal} {Phys. Rev. Lett.}\ }\textbf {\bibinfo {volume} {64}},\ \bibinfo
  {pages} {21} (\bibinfo {year} {1990})}\BibitemShut {NoStop}%
\bibitem [{\citenamefont {Eichler}\ \emph {et~al.}(1986)\citenamefont {Eichler}
  \emph {et~al.}}]{pion1}%
  \BibitemOpen
  \bibfield  {author} {\bibinfo {author} {\bibfnamefont {R.}~\bibnamefont
  {Eichler}} \emph {et~al.},\ }\href@noop {} {\bibfield  {journal} {\bibinfo
  {journal} {Phys. Lett. B}\ }\textbf {\bibinfo {volume} {175}},\ \bibinfo
  {pages} {101} (\bibinfo {year} {1986})}\BibitemShut {NoStop}%
\bibitem [{\citenamefont {Picciotto}\ \emph {et~al.}(1988)\citenamefont
  {Picciotto} \emph {et~al.}}]{pion2}%
  \BibitemOpen
  \bibfield  {author} {\bibinfo {author} {\bibfnamefont {C.~E.}\ \bibnamefont
  {Picciotto}} \emph {et~al.},\ }\href@noop {} {\bibfield  {journal} {\bibinfo
  {journal} {Phys. Rev. D}\ }\textbf {\bibinfo {volume} {37}},\ \bibinfo
  {pages} {1131} (\bibinfo {year} {1988})}\BibitemShut {NoStop}%
\bibitem [{\citenamefont {Aguilar-Arevalo}\ \emph {et~al.}(2021)\citenamefont
  {Aguilar-Arevalo} \emph {et~al.}}]{pion3}%
  \BibitemOpen
  \bibfield  {author} {\bibinfo {author} {\bibfnamefont {A.}~\bibnamefont
  {Aguilar-Arevalo}} \emph {et~al.},\ }\href@noop {} {\bibfield  {journal}
  {\bibinfo  {journal} {Phys. Rev. D}\ }\textbf {\bibinfo {volume} {103}},\
  \bibinfo {pages} {052006} (\bibinfo {year} {2021})}\BibitemShut {NoStop}%
\bibitem [{\citenamefont {Albrecht}\ \emph {et~al.}(1990)\citenamefont
  {Albrecht} \emph {et~al.}}]{tau1}%
  \BibitemOpen
  \bibfield  {author} {\bibinfo {author} {\bibfnamefont {H.}~\bibnamefont
  {Albrecht}} \emph {et~al.},\ }\href@noop {} {\bibfield  {journal} {\bibinfo
  {journal} {Phys. Lett. B}\ }\textbf {\bibinfo {volume} {246}},\ \bibinfo
  {pages} {278} (\bibinfo {year} {1990})}\BibitemShut {NoStop}%
\bibitem [{\citenamefont {Baltrusaitis}\ \emph {et~al.}(1985)\citenamefont
  {Baltrusaitis} \emph {et~al.}}]{tau2}%
  \BibitemOpen
  \bibfield  {author} {\bibinfo {author} {\bibfnamefont {R.~M.}\ \bibnamefont
  {Baltrusaitis}} \emph {et~al.},\ }\href@noop {} {\bibfield  {journal}
  {\bibinfo  {journal} {Phys. Rev. Lett.}\ }\textbf {\bibinfo {volume} {55}},\
  \bibinfo {pages} {1842} (\bibinfo {year} {1985})}\BibitemShut {NoStop}%
\bibitem [{\citenamefont {Derenzo}(1969)}]{search1}%
  \BibitemOpen
  \bibfield  {author} {\bibinfo {author} {\bibfnamefont {S.~E.}\ \bibnamefont
  {Derenzo}},\ }\href@noop {} {\bibfield  {journal} {\bibinfo  {journal} {Phys.
  Rev.}\ }\textbf {\bibinfo {volume} {181}},\ \bibinfo {pages} {1854} (\bibinfo
  {year} {1969})}\BibitemShut {NoStop}%
\bibitem [{\citenamefont {Bryman}\ and\ \citenamefont
  {Clifford}(1986)}]{search2}%
  \BibitemOpen
  \bibfield  {author} {\bibinfo {author} {\bibfnamefont {D.~A.}\ \bibnamefont
  {Bryman}}\ and\ \bibinfo {author} {\bibfnamefont {E.~T.~H.}\ \bibnamefont
  {Clifford}},\ }\href {https://doi.org/10.1103/PhysRevLett.57.2787} {\bibfield
   {journal} {\bibinfo  {journal} {Phys. Rev. Lett.}\ }\textbf {\bibinfo
  {volume} {57}},\ \bibinfo {pages} {2787} (\bibinfo {year}
  {1986})}\BibitemShut {NoStop}%
\bibitem [{\citenamefont {Bilger}\ \emph {et~al.}(1999)\citenamefont {Bilger}
  \emph {et~al.}}]{search3}%
  \BibitemOpen
  \bibfield  {author} {\bibinfo {author} {\bibfnamefont {R.}~\bibnamefont
  {Bilger}} \emph {et~al.},\ }\href@noop {} {\bibfield  {journal} {\bibinfo
  {journal} {Phys. Lett. B}\ }\textbf {\bibinfo {volume} {446}},\ \bibinfo
  {pages} {363 } (\bibinfo {year} {1999})}\BibitemShut {NoStop}%
\bibitem [{\citenamefont {Bayes}\ \emph {et~al.}(2015)\citenamefont {Bayes}
  \emph {et~al.}}]{search4}%
  \BibitemOpen
  \bibfield  {author} {\bibinfo {author} {\bibfnamefont {R.}~\bibnamefont
  {Bayes}} \emph {et~al.},\ }\href@noop {} {\bibfield  {journal} {\bibinfo
  {journal} {Phys. Rev. D}\ }\textbf {\bibinfo {volume} {91}},\ \bibinfo
  {pages} {052020} (\bibinfo {year} {2015})}\BibitemShut {NoStop}%
\bibitem [{\citenamefont {Aguilar-Arevalo}\ \emph {et~al.}(2020)\citenamefont
  {Aguilar-Arevalo} \emph {et~al.}}]{search5}%
  \BibitemOpen
  \bibfield  {author} {\bibinfo {author} {\bibfnamefont {A.}~\bibnamefont
  {Aguilar-Arevalo}} \emph {et~al.},\ }\href@noop {} {\bibfield  {journal}
  {\bibinfo  {journal} {Phys. Rev. D}\ }\textbf {\bibinfo {volume} {101}},\
  \bibinfo {pages} {052014} (\bibinfo {year} {2020})}\BibitemShut {NoStop}%
\bibitem [{\citenamefont {Jodidio}\ \emph {et~al.}(1986)\citenamefont {Jodidio}
  \emph {et~al.}}]{search6}%
  \BibitemOpen
  \bibfield  {author} {\bibinfo {author} {\bibfnamefont {A.}~\bibnamefont
  {Jodidio}} \emph {et~al.},\ }\href@noop {} {\bibfield  {journal} {\bibinfo
  {journal} {Phys. Rev. D}\ }\textbf {\bibinfo {volume} {34}},\ \bibinfo
  {pages} {1967} (\bibinfo {year} {1986})}\BibitemShut {NoStop}%
\bibitem [{\citenamefont {Baldini}\ \emph {et~al.}(2020)\citenamefont {Baldini}
  \emph {et~al.}}]{search7}%
  \BibitemOpen
  \bibfield  {author} {\bibinfo {author} {\bibfnamefont {A.~M.}\ \bibnamefont
  {Baldini}} \emph {et~al.},\ }\href@noop {} {\bibfield  {journal} {\bibinfo
  {journal} {Eur. Phys. J. C}\ }\textbf {\bibinfo {volume} {80}},\ \bibinfo
  {pages} {858} (\bibinfo {year} {2020})}\BibitemShut {NoStop}%
\bibitem [{\citenamefont {Balke}\ \emph {et~al.}(1988)\citenamefont {Balke}
  \emph {et~al.}}]{search8}%
  \BibitemOpen
  \bibfield  {author} {\bibinfo {author} {\bibfnamefont {B.}~\bibnamefont
  {Balke}} \emph {et~al.},\ }\href@noop {} {\bibfield  {journal} {\bibinfo
  {journal} {Phys. Rev. D}\ }\textbf {\bibinfo {volume} {37}},\ \bibinfo
  {pages} {587} (\bibinfo {year} {1988})}\BibitemShut {NoStop}%
\bibitem [{\citenamefont {Heeck}\ and\ \citenamefont
  {Rodejohann}(2018)}]{mot3}%
  \BibitemOpen
  \bibfield  {author} {\bibinfo {author} {\bibfnamefont {J.}~\bibnamefont
  {Heeck}}\ and\ \bibinfo {author} {\bibfnamefont {W.}~\bibnamefont
  {Rodejohann}},\ }\href@noop {} {\bibfield  {journal} {\bibinfo  {journal}
  {Phys. Lett. B}\ }\textbf {\bibinfo {volume} {776}},\ \bibinfo {pages} {385 }
  (\bibinfo {year} {2018})}\BibitemShut {NoStop}%
\bibitem [{\citenamefont {Bernstein}\ and\ \citenamefont
  {Cooper}(2013)}]{clfv1}%
  \BibitemOpen
  \bibfield  {author} {\bibinfo {author} {\bibfnamefont {R.}~\bibnamefont
  {Bernstein}}\ and\ \bibinfo {author} {\bibfnamefont {P.~S.}\ \bibnamefont
  {Cooper}},\ }\href@noop {} {\bibfield  {journal} {\bibinfo  {journal} {Phys.
  Rep.}\ }\textbf {\bibinfo {volume} {532}},\ \bibinfo {pages} {27 } (\bibinfo
  {year} {2013})}\BibitemShut {NoStop}%
\bibitem [{\citenamefont {Michel}(1950)}]{michel1}%
  \BibitemOpen
  \bibfield  {author} {\bibinfo {author} {\bibfnamefont {L.}~\bibnamefont
  {Michel}},\ }\href@noop {} {\bibfield  {journal} {\bibinfo  {journal} {Proc.
  Phys. Soc. A}\ }\textbf {\bibinfo {volume} {63}},\ \bibinfo {pages} {514}
  (\bibinfo {year} {1950})}\BibitemShut {NoStop}%
\bibitem [{\citenamefont {Collar}(2021)}]{myprd}%
  \BibitemOpen
  \bibfield  {author} {\bibinfo {author} {\bibfnamefont {J.~I.}\ \bibnamefont
  {Collar}},\ }\href@noop {} {\bibfield  {journal} {\bibinfo  {journal} {Phys.
  Rev. D}\ }\textbf {\bibinfo {volume} {103}},\ \bibinfo {pages} {052007}
  (\bibinfo {year} {2021})}\BibitemShut {NoStop}%
\bibitem [{\citenamefont {Barbeau}\ \emph {et~al.}(2007)\citenamefont
  {Barbeau}, \citenamefont {Collar},\ and\ \citenamefont {Tench}}]{ppc}%
  \BibitemOpen
  \bibfield  {author} {\bibinfo {author} {\bibfnamefont {P.~S.}\ \bibnamefont
  {Barbeau}}, \bibinfo {author} {\bibfnamefont {J.~I.}\ \bibnamefont
  {Collar}},\ and\ \bibinfo {author} {\bibfnamefont {O.}~\bibnamefont
  {Tench}},\ }\href {https://doi.org/10.1088/1475-7516/2007/09/009} {\bibfield
  {journal} {\bibinfo  {journal} {JCAP}\ }\textbf {\bibinfo {volume}
  {2007}}\bibinfo  {number} { (09)},\ \bibinfo {pages} {009}}\BibitemShut
  {NoStop}%
\bibitem [{\citenamefont {Gninenko}\ and\ \citenamefont
  {Krasnikov}(1998)}]{sergei}%
  \BibitemOpen
\bibfield  {number} {  }\bibfield  {author} {\bibinfo {author} {\bibfnamefont
  {S.}~\bibnamefont {Gninenko}}\ and\ \bibinfo {author} {\bibfnamefont
  {N.}~\bibnamefont {Krasnikov}},\ }\href@noop {} {\bibfield  {journal}
  {\bibinfo  {journal} {Phys. Lett. B}\ }\textbf {\bibinfo {volume} {434}},\
  \bibinfo {pages} {163} (\bibinfo {year} {1998})}\BibitemShut {NoStop}%
\bibitem [{\citenamefont {Beacom}\ and\ \citenamefont {Y\"uksel}(2006)}]{john}%
  \BibitemOpen
  \bibfield  {author} {\bibinfo {author} {\bibfnamefont {J.~F.}\ \bibnamefont
  {Beacom}}\ and\ \bibinfo {author} {\bibfnamefont {H.}~\bibnamefont
  {Y\"uksel}},\ }\href {https://doi.org/10.1103/PhysRevLett.97.071102}
  {\bibfield  {journal} {\bibinfo  {journal} {Phys. Rev. Lett.}\ }\textbf
  {\bibinfo {volume} {97}},\ \bibinfo {pages} {071102} (\bibinfo {year}
  {2006})}\BibitemShut {NoStop}%
\bibitem [{\citenamefont {Sizun}\ \emph {et~al.}(2006)\citenamefont {Sizun},
  \citenamefont {Cass\'e},\ and\ \citenamefont {Schanne}}]{john2}%
  \BibitemOpen
  \bibfield  {author} {\bibinfo {author} {\bibfnamefont {P.}~\bibnamefont
  {Sizun}}, \bibinfo {author} {\bibfnamefont {M.}~\bibnamefont {Cass\'e}},\
  and\ \bibinfo {author} {\bibfnamefont {S.}~\bibnamefont {Schanne}},\ }\href
  {https://doi.org/10.1103/PhysRevD.74.063514} {\bibfield  {journal} {\bibinfo
  {journal} {Phys. Rev. D}\ }\textbf {\bibinfo {volume} {74}},\ \bibinfo
  {pages} {063514} (\bibinfo {year} {2006})}\BibitemShut {NoStop}%
\bibitem [{\citenamefont {Siegert}(2023)}]{5113}%
  \BibitemOpen
  \bibfield  {author} {\bibinfo {author} {\bibfnamefont {T.}~\bibnamefont
  {Siegert}},\ }\href@noop {} {\bibfield  {journal} {\bibinfo  {journal}
  {Astrophys. Space Sci.}\ }\textbf {\bibinfo {volume} {368}},\ \bibinfo
  {pages} {27} (\bibinfo {year} {2023})}\BibitemShut {NoStop}%
\bibitem [{\citenamefont {Prantzos}\ \emph {et~al.}(2011)\citenamefont
  {Prantzos} \emph {et~al.}}]{5111}%
  \BibitemOpen
  \bibfield  {author} {\bibinfo {author} {\bibfnamefont {N.}~\bibnamefont
  {Prantzos}} \emph {et~al.},\ }\href@noop {} {\bibfield  {journal} {\bibinfo
  {journal} {Rev. Mod. Phys.}\ }\textbf {\bibinfo {volume} {83}},\ \bibinfo
  {pages} {1001} (\bibinfo {year} {2011})}\BibitemShut {NoStop}%
\bibitem [{\citenamefont {Kierans}\ \emph {et~al.}()\citenamefont {Kierans}
  \emph {et~al.}}]{5112}%
  \BibitemOpen
  \bibfield  {author} {\bibinfo {author} {\bibfnamefont {C.~A.}\ \bibnamefont
  {Kierans}} \emph {et~al.},\ }\href@noop {} {}\Eprint
  {https://arxiv.org/abs/arXiv:1903.05569} {arXiv:1903.05569} \BibitemShut
  {NoStop}%
\bibitem [{leg()}]{lege}%
  \BibitemOpen
  \href@noop {} {}\bibinfo {note} {Low Energy Germanium (LEGe), commercially
  available from Mirion Technologies Inc.}\BibitemShut {Stop}%
\bibitem [{\citenamefont {Pifer}\ \emph {et~al.}(1976)\citenamefont {Pifer},
  \citenamefont {Bowen},\ and\ \citenamefont {Kendall}}]{surface}%
  \BibitemOpen
  \bibfield  {author} {\bibinfo {author} {\bibfnamefont {A.}~\bibnamefont
  {Pifer}}, \bibinfo {author} {\bibfnamefont {T.}~\bibnamefont {Bowen}},\ and\
  \bibinfo {author} {\bibfnamefont {K.}~\bibnamefont {Kendall}},\ }\href@noop
  {} {\bibfield  {journal} {\bibinfo  {journal} {Nucl. Instr. Meth.}\ }\textbf
  {\bibinfo {volume} {135}},\ \bibinfo {pages} {39} (\bibinfo {year}
  {1976})}\BibitemShut {NoStop}%
\bibitem [{\citenamefont {Marshall}(1992)}]{triumf}%
  \BibitemOpen
  \bibfield  {author} {\bibinfo {author} {\bibfnamefont {G.~M.}\ \bibnamefont
  {Marshall}},\ }\href@noop {} {\bibfield  {journal} {\bibinfo  {journal} {Z.
  Phys. C}\ }\textbf {\bibinfo {volume} {56}},\ \bibinfo {pages} {226}
  (\bibinfo {year} {1992})}\BibitemShut {NoStop}%
\bibitem [{elj()}]{eljen}%
  \BibitemOpen
  \href@noop {} {}\bibinfo {note} {EJ-214 available from Eljen Technology,
  Sweetwater, TX}\BibitemShut {NoStop}%
\bibitem [{\citenamefont {Lewis}(2023)}]{mark}%
  \BibitemOpen
  \bibfield  {author} {\bibinfo {author} {\bibfnamefont {C.~M.}\ \bibnamefont
  {Lewis}},\ }\href@noop {} {Ph.D. thesis},\ \bibinfo  {school} {University of
  Chicago} (\bibinfo {year} {2023})\BibitemShut {NoStop}%
\bibitem [{\citenamefont {Colaresi}\ \emph {et~al.}(2021)\citenamefont
  {Colaresi} \emph {et~al.}}]{dresden1}%
  \BibitemOpen
  \bibfield  {author} {\bibinfo {author} {\bibfnamefont {J.}~\bibnamefont
  {Colaresi}} \emph {et~al.},\ }\href@noop {} {\bibfield  {journal} {\bibinfo
  {journal} {Phys. Rev. D}\ }\textbf {\bibinfo {volume} {104}},\ \bibinfo
  {pages} {072003} (\bibinfo {year} {2021})}\BibitemShut {NoStop}%
\bibitem [{\citenamefont {Colaresi}\ \emph {et~al.}(2022)\citenamefont
  {Colaresi}, \citenamefont {Collar}, \citenamefont {Hossbach}, \citenamefont
  {Lewis},\ and\ \citenamefont {Yocum}}]{dresden2}%
  \BibitemOpen
  \bibfield  {author} {\bibinfo {author} {\bibfnamefont {J.}~\bibnamefont
  {Colaresi}}, \bibinfo {author} {\bibfnamefont {J.~I.}\ \bibnamefont
  {Collar}}, \bibinfo {author} {\bibfnamefont {T.~W.}\ \bibnamefont
  {Hossbach}}, \bibinfo {author} {\bibfnamefont {C.~M.}\ \bibnamefont
  {Lewis}},\ and\ \bibinfo {author} {\bibfnamefont {K.~M.}\ \bibnamefont
  {Yocum}},\ }\href@noop {} {\bibfield  {journal} {\bibinfo  {journal} {Phys.
  Rev. Lett.}\ }\textbf {\bibinfo {volume} {129}},\ \bibinfo {pages} {211802}
  (\bibinfo {year} {2022})}\BibitemShut {NoStop}%
\bibitem [{csd()}]{csda1}%
  \BibitemOpen
  \href@noop {} {}\bibinfo {note}
  {\url{https://physics.nist.gov/PhysRefData/Star/Text/ESTAR.html}}\BibitemShut
  {NoStop}%
\bibitem [{\citenamefont {Batra}\ and\ \citenamefont {Sehgal}(1981)}]{csda2}%
  \BibitemOpen
  \bibfield  {author} {\bibinfo {author} {\bibfnamefont {R.~K.}\ \bibnamefont
  {Batra}}\ and\ \bibinfo {author} {\bibfnamefont {M.~L.}\ \bibnamefont
  {Sehgal}},\ }\href {https://doi.org/10.1103/PhysRevB.23.4448} {\bibfield
  {journal} {\bibinfo  {journal} {Phys. Rev. B}\ }\textbf {\bibinfo {volume}
  {23}},\ \bibinfo {pages} {4448} (\bibinfo {year} {1981})}\BibitemShut
  {NoStop}%
\bibitem [{mcn()}]{mcnpx}%
  \BibitemOpen
  \href@noop {} {}\bibinfo {note} {D.B. Pelowitz {\it et al.}, Los Alamos
  National Laboratory report LA-UR-11-02295 (2011).}\BibitemShut {Stop}%
\bibitem [{\citenamefont {Hesketh}\ \emph {et~al.}(2022)\citenamefont {Hesketh}
  \emph {et~al.}}]{mu3e}%
  \BibitemOpen
  \bibfield  {author} {\bibinfo {author} {\bibfnamefont {G.}~\bibnamefont
  {Hesketh}} \emph {et~al.} (\bibinfo {collaboration} {Mu3e}),\ }in\ \href@noop
  {} {\emph {\bibinfo {booktitle} {{Snowmass 2021}}}}\ (\bibinfo {year}
  {2022})\ \Eprint {https://arxiv.org/abs/2204.00001} {arXiv:2204.00001}
  \BibitemShut {NoStop}%
\bibitem [{\citenamefont {{Luke}}\ \emph {et~al.}(1989)\citenamefont {{Luke}},
  \citenamefont {{Goulding}}, \citenamefont {{Madden}},\ and\ \citenamefont
  {{Pehl}}}]{Luke}%
  \BibitemOpen
  \bibfield  {author} {\bibinfo {author} {\bibfnamefont {P.~N.}\ \bibnamefont
  {{Luke}}}, \bibinfo {author} {\bibfnamefont {F.~S.}\ \bibnamefont
  {{Goulding}}}, \bibinfo {author} {\bibfnamefont {N.~W.}\ \bibnamefont
  {{Madden}}},\ and\ \bibinfo {author} {\bibfnamefont {R.~H.}\ \bibnamefont
  {{Pehl}}},\ }\href@noop {} {\bibfield  {journal} {\bibinfo  {journal} {IEEE
  Trans. Nucl. Sci.}\ }\textbf {\bibinfo {volume} {36}},\ \bibinfo {pages}
  {926} (\bibinfo {year} {1989})}\BibitemShut {NoStop}%
\end{thebibliography}%

\end{document}